\documentclass[twocolumn,amsmath,amssymb,prd]{revtex4}

\def\al{\alpha}
\def\be{\beta}
\def\ga{\gamma}
\def\de{\delta}

\def\th{\theta}

\def\la{\lambda}

\def\ph{\phi}

\def\ch{\chi}

\def\om{\omega}
\def\Ga{\Gamma}
\def\De{\Delta}

\def\Om{\Omega}

\def\lsim{\mathrel{\rlap{\lower4pt\hbox{\hskip1pt$\sim$}}
    \raise1pt\hbox{$<$}}}
\def\gsim{\mathrel{\rlap{\lower4pt\hbox{\hskip1pt$\sim$}}
    \raise1pt\hbox{$>$}}}
\def\Re{\hbox{Re}\,}
\def\Im{\hbox{Im}\,}

\def\etal {{\it et al.}}
\newcommand{\beq}{\begin{equation}}
\newcommand{\eeq}{\end{equation}}
\newcommand{\bea}{\begin{eqnarray}}
\newcommand{\eea}{\end{eqnarray}}
\newcommand{\bse}{\begin{subequations}}
\newcommand{\ese}{\end{subequations}}

\def\ni{\noindent}
\def\sF#1#2{{\textstyle{{#1}\over{#2}}\,}}
\newcommand{\BB}{\big}
\newcommand{\nn}{\nonumber}

\def\to{\rightarrow}

\def\nub{\bar\nu}

\def\nub{\bar\nu}

\def\C#1{({\cal C})_{#1}}

\def\ol#1{\overline{#1}}

\usepackage{amsmath}
\usepackage{amsfonts}
\usepackage{amssymb}

\usepackage{float}
\usepackage{graphicx}
\usepackage[usenames,dvipsnames]{color}
\usepackage{wrapfig}

\usepackage{slashed}

\newcommand{\bM}{\begin{pmatrix}}
\newcommand{\eM}{\end{pmatrix}}
\newcommand{\W}{\widetilde}

\def\ring#1{{\mathaccent'27 #1}}

\def\ari{\ring{a}}

\newcommand{\n}{\hat{\pmb{n}}}
\newcommand{\ha}{\sF{1}{2}}

\newcommand{\of}{\text{of}}
\newcommand{\eff}{\text{eff}}
\newcommand{\Aof}{a^{(3)}_{\of}}
\newcommand{\arit}{\ari^{(3)}_{\of}}

\newcommand{\Bz}{(B_\text{exp})_0}
\newcommand{\wT}{\om_\oplus T_\oplus}

\def\A#1#2#3#4{(a^{(#1)}_{#2})_{#3}^{#4}}
\def\C#1#2#3{(c^{(#1)}_{#2})_{#3}}

\begin{document}

\title{Tests of Lorentz symmetry in single beta decay}
\author{Jorge S. D\'iaz}
\affiliation{Institute for Theoretical Physics, Karlsruhe Institute of Technology, 76128 Karlsruhe, Germany}

\begin{abstract}

Low-energy experiments studying single beta decay can serve as sensitive probes of Lorentz invariance that can complement interferometric searches for deviations from this spacetime symmetry.
Experimental signatures of a dimension-three operator for Lorentz violation that are unobservable in neutrino oscillations are described for the decay the polarized and unpolarized neutrons as well as for measurements of the spectral endpoint in beta decay.

\end{abstract}

\maketitle

\section{Introduction}

The foundations of modern physics assume the invariance of physical laws under rotations and boosts, known as Lorentz symmetry.
In our search for new physics, the possibility of minute violations of Lorentz invariance has become an active field of study by the development of theoretical formalisms and mainly by searching for key signatures in a wide range of experiments \cite{tables}.
Precise studies of beta decay offer the opportunity to search for physics beyond the Standard Model. For instance, many experiments measuring the decay of neutrons are searching for unconventional couplings in weak interactions leading to new sources of CP violation \cite{NicoSnow}.
Similarly, the search for a distortion in the spectrum of tritium decay would provide an absolute measurement of the neutrino mass.
These experiments can also search for deviations from exact Lorentz symmetry.
The interferometric nature of quantum oscillations give neutral mesons \cite{meson_osc} and neutrinos \cite{LVnu} a remarkable sensitivity to signals of new physics; nonetheless, there are certain signals than are unobservable in these experiments.
For neutrinos, it has been shown that beta-decay experiments have unique sensitivity to so-called {\it countershaded} operators, which produce no effects in oscillations nor modifications of the neutrino velocity; therefore, their effects can only be studied via weak decays \cite{DKL}.

This paper describes the relevant signatures of Lorentz and CPT violation in single beta decay experiments.
High-precision measurements of beta-decay spectra for the determination of neutron and neutrino properties offer an attractive opportunity to test Lorentz invariance by searching for distinctive signals that could arise in current and future experiments.
Observable effects can also appear in double beta decays \cite{Diaz_DBD}.
Systematic searches for Lorentz violation in experiments use a general framework based on effective field theory known as the Standard-Model Extension (SME) \cite{SME1a,SME1b,SME2}. 
This framework incorporates coordinate-independent terms that break Lorentz symmetry in the Standard Model action in the form of conventional operators contracted with controlling coefficients for Lorentz violation.
These terms can trigger observable signals under the rotation and/or boost of the relevant experimental system.
The spontaneous breakdown of Lorentz symmetry at high energies in some string-theory scenarios \cite{SBS_LV1} suggests that the SME coefficients should be small due to the relevant energy scale suppression, such as the Planck scale.
Nevertheless, potentially large deviations from Lorentz symmetry have been considered in systems involving weak decays \cite{DKL} and matter-gravity couplings \cite{KT}.

For neutrinos, the SME has been used to search for signatures of Lorentz violation in oscillations \cite{KM_SB,DKM} using accelerators \cite{LV_LSND,LV_MiniBooNE1,LV_MiniBooNE2,LV_MINOS_ND1,LV_MINOS_ND2,LV_MINOS_FD,RebelMufson}, atmospheric neutrinos \cite{LV_IceCube,LV_SK}, and reactors \cite{LV_DC,LV_DC2}, reaching impressive sensitivity.
Neutrino oscillations are powerful tools to test Lorentz symmetry;
nonetheless, there are operators that are unobservable in these type of experiments.
For these oscillation-free operators, other kinematical effects must be invoked such as modifications of the neutrino velocity, which lead to Cherenkov radiation and threshold effects \cite{Diaz_LVreview}.
In these scenarios, the effects of Lorentz violation can be enhanced by the neutrino energy and propagation time, which makes astrophysical sources sensitive probes of Lorentz symmetry \cite{DKM2}, particularly when operators of arbitrary dimension are incorporated in the action \cite{KM2012}.

The fundamental role of beta-decay experiments is the study of countershaded operators, that are oscillation-free terms of mass dimension three in the Lagrangian that are controlled by the SME coefficient $\Aof$.
These CPT-odd operators leave the neutrino velocity unchanged and their experimental signatures are unaffected by the neutrino energy \cite{KM2012}.
This feature makes beta-decay experiments unique probes of Lorentz symmetry \cite{DKL}. 
For illustration purposes, some neutron experiments such as $a$CORN \cite{aCORN}, $a$SPECT \cite{aSPECT}, and PERKEO \cite{PERKEO} are discussed; nevertheless, several of the observable signatures can be studied by other experiments including 
$ab$BA \cite{abBA}, 
emiT \cite{emiT},
N$ab$ \cite{Nab}, 
nTRV \cite{nTRV},
PERC \cite{PERC}, and
UCNB \cite{UCNB}.
Similarly, the analysis of tritium decay can be applied to experiments for neutrino-mass measurements 
Mainz \cite{Mainz}, 
Troitsk \cite{Troitsk}, and 
Karlsruhe Tritium Neutrino experiment (KATRIN) \cite{KATRIN} 
as well as the 
Princeton Tritium Observatory for Light, Early-Universe, Massive-Neutrino Yield (PTOLEMY) \cite{PTOLEMY}
proposed to search for the cosmic neutrino background.

\section{Beta decay}

At low energies, the transition amplitude describing nuclear beta decay is well described by the Fermi four-fermion interaction of the form
\beq\label{M}
i\mathcal{M} = \frac{iG_F}{\sqrt2}V_{ud}
\BB[\ol u(p)\ga_\al(1-\ga_5)v(q)\BB]
J^\al,
\eeq
where $J^\al$ is the current describing the nuclear states, the spinor $\ol u(p)$ corresponds to the emitted electron of 4-momentum $p^\al=(E,\pmb{p})$ and mass $m_e$, and the antineutrino of mass $m_\nu$ and 4-momentum $q^\al=(\om,\pmb{q})$ is given by spinor $v(q)$.
The constant factors are the Fermi constant $G_F$ and the relevant element of the CKM matrix $V_{ud}\approx\cos\theta_C$.
In this work we are interested in the potential breakdown of Lorentz invariance in the neutrino sector \cite{Diaz_LVreview}.
Recent studies have also considered Lorentz-violating effects in weak decays arising in the gauge sector \cite{NoordmansPRC,Altschul}.
The emitted antineutrino escapes unmeasured in beta decay experiments; however, imprints of its behavior can be inferred from the decay products experimentally accessible.
The effects of Lorentz violation are controlled by the four components of the coefficient $\Aof$, which in spherical basis introduced in Ref. \cite{KM2012} are denoted by $(\Aof)_{00}$, $(\Aof)_{10}$, $\Re(\Aof)_{11}$, and $\Im(\Aof)_{11}$.
To date, only theoretically estimated bounds exist on $(\Aof)_{00}$ and $(\Aof)_{10}$ \cite{tables,DKL,KM2012}, whereas the real and imaginary parts of $(\Aof)_{11}$ remain unexplored.
Below we present the signatures of these four coefficients so they can be directly studied in experiments.

\section{Neutron decay}

For the transition amplitude in Eq. \eqref{M} describing beta decay of a neutron we write the nuclear current in the form
\beq
J^\al = \ol u_p\ga^\al(1+\la\ga_5)u_n,
\eeq
where $u_n$ and $\ol u_p$ represent the neutron and proton, respectively, and $\la=g_A/g_V$ is the ratio between the axial and vector couplings.
It is important to emphasize that Lorentz violation modifies the neutrino dispersion relations and the spinor solutions satisfy a modified equation of motion.

The sum over the final spin states allowing for a polarized neutron in the direction $\n$ can be written in terms of the electron energy $E$ and velocity $\pmb{\be}$ in the form \cite{Jackson57}
\beq
\sum_\text{spin}|\mathcal{M}|^2 = 16M^2_nC\, E\om
\BB(1+a\,\pmb{\be}\cdot\hat{\W{\pmb{q}}}
+A\,\n\cdot\pmb{\be}
+B\,\n\cdot\hat{\W{\pmb{q}}}
\BB),
\label{|M|^2_neutron}
\eeq

\ni
where Lorentz-violating effects appear in the form of an effective momentum $\hat{\W{\pmb{q}}}=(\pmb{q}+\pmb{a}_\text{of}^{(3)}-\arit\,\hat{\pmb{q}})/\om$ for the antineutrino, with the isotropic component of $\Aof$ denoted by $\arit=(\Aof)_{00}/\sqrt{4\pi}$.
The constant factor is given by $C=G_F^2|V_{ud}|^2\,(1+3\la^2)$, the nucleon mass is $M_n$, and the correlation parameters are given by the conventional definitions \cite{NicoSnow}
\beq\label{aAB}
a = \frac{1-\la^2}{1+3\la^2},\;
A =-\frac{2\la(\la+1)}{1+3\la^2},\;
B = 2\frac{\la(\la-1)}{1+3\la^2}.
\eeq

The decay rate is given by
\bea\label{dGamma}
d\Ga 
&=& \frac{1}{4M^2_n} \int
\frac{d^3p}{(2\pi)^32E}
\frac{d^3q}{(2\pi)^32\om} \,
F(Z,E)
\nn\\
&&\quad\quad\times
\sum_\text{spin}|\mathcal{M}|^2
2\pi\delta(E_A-E_B-E-\om), 
\eea
where the Fermi function has been included to incorporate the electrostatic interaction between the proton ($Z=1$) and the outgoing electron.
Integrating over the antineutrino energy $\om$ and using $d^3p = |\pmb{p}|E \, dE d\Om_e$, $d^3q = (\om^2+2\om\arit) \, d\om d\Om_{\nub}$ we can write the electron differential spectrum
\bea\label{dGamma2}
\frac{d\Ga}{d\Om_e\,d\Om_{\nub}\,dT} &=& \frac{C}{(2\pi)^5} 
F(Z,E)|\pmb{p}| E\, 
(\om^2_0+2\om_0\arit)
\nn\\
&&\times
\Big(1 + a\, \pmb{\be}\cdot \hat{\W{\pmb{q}}} 
+
A\,\n\cdot\pmb{\be}+ B\,\n\cdot\hat{\W{\pmb{q}}}\Big)
,
\eea
with $\om_0=T_0-T$. The kinetic energy of the electron is given by $T=E-m_e$ and $T_0$ denotes the maximum kinetic energy available in the decay.

\subsection{Unpolarized neutrons}

Experiments with unpolarized neutrons ($\n=0$) can be classified into two categories: those that only measure the electron spectrum and those in which the relative orientation between the two emitted leptons can be identified, relevant for the measurement of the electron-antineutrino asymmetry $a$ defined in Eq. \eqref{aAB}.
The signatures of Lorentz violation for these two cases are presented below.

\subsubsection{Spectrum measurements}

For experiments with unpolarized neutrons that only measure the outgoing electrons, the directions of the two final leptons can be integrated to reduce the spectrum \eqref{dGamma2} to the form
\bea
\frac{d\Ga}{dT} &=& \frac{C}{2\pi^3} \,F(Z,T)
\,|\pmb{p}| (T+m_e) \, 
\BB(\om_0^2 + 2\arit\om_0\BB).
\eea

\ni
This expression shows that the Lorentz-violating modification to the antineutrino spinors \eqref{|M|^2_neutron} plays no role and the isotropic effect is purely kinematic due to the modified antineutrino phase space.
The effect of Lorentz violation appears as a small perturbation of the beta-decay spectrum, similar to the effect in two-neutrino double beta decay \cite{Diaz_DBD}.
The energy dependence of the modification introduced by the isotropic coefficient $\arit$ can be used to determine the relevant energy for studying this modification; thus, serving as a guide for a future experimental search of this effect.

The exact value requires a numerical solution due to the involved energy dependence of the Fermi function; nevertheless, a reasonable estimate can be obtained by considering that for neutron decay this function remains almost constant for energies above 200 keV. 
Using this approximation, the maximum of the Lorentz-invariant spectrum satisfies the cubic equation
\beq\label{Tmax_LI}
0=4T^3+2(5m_e-T_0)T^2+m_e(5m_e-4T_0)T-m_e^2T_0.
\eeq
Using $T_0=780$ keV and $m_e=511$ keV, we find that in the absence of Lorentz violation the spectrum has its maximum at $246$ keV.
In the presence of Lorentz violation the maximum gets shifted. 
Instead of the cubic equation \eqref{Tmax_LI}, the maximum of the modified spectrum satisfies the quartic equation
\bea\label{Tmax_LV}
0&=&4T^4+2\BB(5m_e-3T_0-3\arit\BB)T^3
\nn\\
&&
+\BB(2T_0^2+5m_e^2-14m_eT_0+4\arit T_0-14m_e\arit\BB)T^2
\nn\\
&&
+2m_e\BB(2T_0^2-3m_eT_0-3m_e\arit+4\arit T_0\BB)T
\nn\\
&&
+\BB(2m_e^2\arit T_0+m_e^2T_0^2\BB).
\eea
Here we find a method to search for a nonzero value of the coefficient $\arit$: the maximum of the spectrum can be experimentally determined and its value can be replaced in equation \eqref{Tmax_LV}, which gives a linear equation for $\arit$.

It should be noticed that this shift in the maximum of the spectrum can be small and the application of the method mentioned above will depend on the resolution of the experiment.
An equivalent method is the search for a direct deviation of the experimental spectrum from the expected prediction in the absence of Lorentz violation.
This deviation or {\it residual spectrum} reaches its maximum at a well defined energy $T_m$ which is independent on the size of $\arit$ and satisfies the cubic equation
\bea
0 &=& 3T_m^3 + (7m_e-2T_0)T_m^2 
\nn\\
&&\qquad
+ \, m_e(3m_e-4T_0)T_m - m_e^2T_0.
\eea
For the numerical values used before, we find that the effect of a nonzero coefficient $\arit$ is maximal at $T_m=406$ keV; hence, this is the region of the spectrum where deviations from the conventional spectrum should be explored. 
The size of the deviation corresponds to a direct measurement of the magnitude of the coefficient $\arit$.

\subsubsection{Electron-antineutrino asymmetry}

The decay of unpolarized neutrons is also used to measure the antineutrino-electron asymmetry $a$ in Eq. \eqref{|M|^2_neutron}. 
The $a$CORN experiment has a proton detector and an electron detector aligned perpendicular to the neutron beam, in which only decay products emitted along the axis of the detectors are analyzed \cite{aCORN}. 
The design of the experiment allows identifying events in which the two leptons are emitted parallel $N_+$ and antiparallel $N_-$, which can be determined from the spectrum and time-of-flight measurements. 
Since the directionality of the emitted antineutrino can be inferred from the measurement, we have access to the anisotropic coefficients $(\Aof)_{1m}$, with $m=0,\pm1$.

Let us define the number of events in a given direction as
\bea\label{N(T)}
N(T) &=& \frac{d\Ga}{dT\,d\Om_e\,d\Om_{\nub}}
\nn\\
&=&
\W C(T)\, 
\BB(1 + a\, \pmb{\be}\cdot \hat{\W{\pmb{q}}} \BB)
\BB(1+2\Aof/\om_0\BB),
\eea

\ni
where we have defined the function $\W C(T)=C\,F(Z,T)
\,|\pmb{p}| E\, \om_0^2/(2\pi)^5$ and the coefficient $\Aof$ involves isotropic $(\Aof)_{00}$ and anisotropic components $(\Aof)_{1m}$.
The experimental asymmetry $a_\text{exp}$ is defined as
\beq\label{a_exp}
a_\text{exp} =
\frac{N_+-N_-}{N_++N_-},
\eeq
which provides a direct measure of the parameter $a$ defined in Eq. \eqref{aAB}.
From the number of events in a determined orientation given by Eq. \eqref{N(T)}, we find that at leading order the experimental asymmetry \eqref{a_exp} becomes
\bea\label{a_exp_LV} 
a_\text{exp}
&=&
a|\pmb\be|
+\sqrt{\frac{3}{\pi}}\frac{(a^2\pmb\be^2-1)}{\om_0}
\A{3}{\of}{10}{\text{lab}},
\eea
where the first term corresponds to the conventional expression for determining the parameter $a$, with the electron speed given in terms of its kinetic energy as $|\pmb\be|=\sqrt{T(T+2m_e)}/(T+m_e)$.
The second term in Eq. \eqref{a_exp_LV} corresponds to the Lorentz-violating part written in the laboratory frame.
Direct comparison between different experimental searches for Lorentz violation in a physically meaningful way requires a common reference frame, for which the Sun-centered frame is widely used in the literature for reporting constraints on SME coefficients \cite{tables}.
The transformation to this frame is obtained by a sequence of rotations of the form \cite{KM2012}
\beq\label{a10(lab)2a(Sun)}
(\Aof)_{10}^\text{lab} =
\sum_m e^{im\wT} 
d^{(1)}_{0m}(-\ch)\,(\Aof)_{1m},
\eeq
where $d^{(1)}_{0m}(-\ch)$ are the little Wigner matrices and $\ch$ is the colatitude of the experiment.
The dependence on the sidereal time $T_\oplus$ is a consequence of the variation of the coupling between the SME coefficient and the antineutrino direction of propagation due to the Earth rotation with frequency $\om_\oplus\simeq2\pi/(\text{23 h 56 min})$.
The explicit form of expression \eqref{a10(lab)2a(Sun)} is
\bea\label{a1m(Sun)}
\A{3}{\of}{10}{\text{lab}}&=&
\cos\ch\, \A{3}{\of}{10}
\nn\\
&&
+\,\sqrt{2}\, \sin\ch\,
\Im\A{3}{\of}{11}\, \sin\om_\oplus T_\oplus 
\nn\\
&&
-\,\sqrt{2}\, \sin\ch\,\Re\A{3}{\of}{11}\, \cos\om_\oplus T_\oplus. 
\eea
Equation \eqref{a_exp_LV} shows that the effect of Lorentz violation becomes more noticeable near the endpoint; however, for electron energies above 400 keV the measurement of the asymmetry becomes hard because the low energy of the protons makes difficult the proper identification of $N_+$ and $N_-$ using the proton time-of-flight method.
In order to properly measure the asymmetry, the beta spectrometer runs in the range 50 to 350 keV \cite{aCORN}. 

Another experiment designed to measure the parameter $a$ is $a$SPECT, in which a magnetic field perpendicular to the neutron beam guides the protons emitted in the decay towards a proton detector for a precise measurement of the proton spectrum \cite{aSPECT}.
Protons emitted in the opposite direction of the detector are reflected by an electrostatic mirror; thus, the detector can examine protons emitted in all directions.
This feature of the experimental setup makes $a$SPECT insensitive to the Lorentz-violating anisotropies produced by the coefficients $\A{3}{\of}{1m}{}$.
Nonetheless, data collected with the electrostatic mirror switched off allowing only a $2\pi$ coverage can be used to implement a search for anisotropies \cite{WHeil}.

\subsection{Polarized neutrons}

Experiments with polarized neutrons that measure both the beta electron and the recoiling proton can reconstruct the direction of the emitted antineutrino. 
Experiments such as $ab$BA \cite{abBA}, emiT \cite{emiT}, PERC \cite{PERC}, PERKEO \cite{PERKEO}, and UCNB \cite{UCNB} could access anisotropic effects due to Lorentz violation. For instance, unconventional energy- and direction-dependent effects could be studied by an experimental setup for the measurement of the spin-antineutrino asymmetry parameter $B$ in Eq. \eqref{|M|^2_neutron}.

For the decay of polarized neutrons, Lorentz-violating effects appear due to the modified spinor solutions as well as the unconventional antineutrino phase space.
Although the antineutrino escapes unmeasured, the direction of its momentum can be inferred if both the electron and the proton are emitted in the same direction because conservation of momentum along the neutron spin axis can be used to write
\bea
0&=&
\n\cdot\pmb{q}+
\n\cdot\pmb{p}+
\n\cdot\pmb{k}
\nn\\
&=& |\pmb{q}|\cos\th_{\nub}+|\pmb{p}|\cos\th_e+|\pmb{k}|\cos\th_p.
\label{ConsP}
\eea
For this reason, an asymmetry for coincident events in which both the electron and the proton are emitted in the same direction is appropriate for the determination of the parameter $B$ that appears with the antineutrino momentum \cite{Gluck}. 

The number of events in which the electron and the proton are emitted along the direction of the neutron spin is $N^{++}=Q^{++}\W C(T)$, where 
\bea\label{Q++}
Q^{++} 
&=&
\int_{\Om_e^+}d\Om_e \int_{\Om_{\nub}^-} d\Om_{\nub}\,
\BB(1+2\Aof/\om_0\BB)
\nn\\
&&\qquad
\times\,(1
+ a\,\pmb{\be}\cdot \hat{\W{\pmb{q}}}
+ A\,\n\cdot\pmb{\be}
+ B\,\n\cdot\hat{\W{\pmb{q}}}).
\eea
The integration range for the electron and the antineutrino are related by the constraint \eqref{ConsP}, which implies that when the proton is emitted perpendicular to the neutron spin then the antineutrino polar angle $\th_{\nub}$ can take the maximum value $\cos\th_{\nub}=-r\cos\th_e$, with $r=\sqrt{T(T+2m_e)}/(T_0-T)$. 
The integration regions are given by $\Om_{\nub}^-:\,\phi_{\nub}\in[0,2\pi],\,\cos\th_{\nub}\in[-1,-r\cos\th_e]$, $\Om_e^+:\,\phi_e\in[0,2\pi],\,\cos\th_e\in[0,\cos\th_e^\text{max}]$, 
where we have defined $\cos\th_e^\text{max}=1\,(r^{-1})$ for $r<1$ ($r>1$).
The number of events in which the electron and the proton are emitted against the direction of the neutron spin $N^{--}=Q^{--}\W C(T)$ is found directly from $N^{++}$ by reversing the sign of the parameters $A$ and $B$.
We can now define the experimental asymmetry
\beq\label{B_exp}
B_\text{exp} 
= \frac{N^{--}-N^{++}}{N^{--}+N^{++}}
= \frac{Q^{--}-Q^{++}}{Q^{--}+Q^{++}}.
\eeq
Depending on the range of the parameter $r$ and keeping leading-order terms, the experimental asymmetry can be written in the form
$B_\text{exp} = (B_\text{exp})_0 + \de B_\text{exp}$, where the conventional asymmetry takes the form \cite{Gluck}
\beq
(B_\text{exp})_0 =
\frac{4}{3}\left\{ \begin{array}{ccc}
\dfrac{[A\be(2r-3)+B(3-r^2)]}{8-4r+a\be(r^2-2)} &,& r<1 \\\\
\dfrac{-A\be+2Br}{4r-a\be} 
&,& r>1 \end{array}\right.,
\eeq
and the Lorentz-violating modification can be written as
\bea\label{B(wT)}
\de B_\text{exp} &=& \de B_\mathcal{C} + \de B_{\mathcal{A}_s}\sin\wT 
\nn\\
&&\quad
+\, \de B_{\mathcal{A}_c}\cos\wT,
\eea
which explicitly shows the sidereal-time dependence of this quantity.
The amplitudes $\de B_\mathcal{C}$, $\de B_{\mathcal{A}_s}$, $\de B_{\mathcal{A}_c}$ are functions of the location of the apparatus and the electron energy.
They are explicitly presented in Appendix \ref{App_B}.

\section{Spectrum endpoint measurements}

Direct measurements of the neutrino mass $m_\nu$ can be performed by searching for a spectral distortion near the endpoint of beta decay, for which tritium appears as an ideal isotope \cite{Konopinski}.
In an isotropic decay, the anisotropies produced by Lorentz violation are usually unobservable; nonetheless, the use of inhomogeneous magnetic fields for guiding electrons into electrostatic filters (MAC-E) allows selecting electrons emitted in determined directions.
Superconducting magnets produce the guiding magnetic field for the electrons isotropically emitted from the decay of gaseous tritium. 
Electrons with very long paths within the tritium source exhibit a high scattering probability; therefore, only electrons with short paths are accepted to be analyzed. Varying the magnetic field from a value in the tritium source $B_\text{S}$ to a maximum value $B_\text{max}$ creates a cone of acceptance of aperture $\theta_0$, with 
\beq\label{theta_acc}
\sin\theta_0 = \sqrt\frac{B_\text{S}}{B_\text{max}}.
\eeq
Electrons emitted at angles $\th>\th_0$ are reflected due to a magnetic mirror effect.
This selection is what permits the study of anisotropic effects.

Given the configuration of tritium-decay experiments, the sequence of rotations implemented for relating the components of $\Aof$ in the laboratory frame to the relevant components in the Sun-centered frame differs from the one used in the previous section and it takes the explicit form \cite{DKL}
\bea\label{a3_tritium}
(\Aof)^\text{lab} &=& \sum_{jm}e^{im\wT}\sum_{m'm''} Y_{jm'}(\th,\ph) \, d^{(j)}_{m'm''}(-\pi/2)
\nn\\
&&\quad\times\;
e^{-im''\xi} \, d^{(j)}_{m''m}(-\ch) \, (\Aof)_{jm},
\eea
where the extra rotations implemented in this transformation set the laboratory $z$ axis along the direction of the axis of the experiment determined by the magnetic field in the decay region. 
The spherical harmonics $Y_{jm'}(\th,\ph)$ are written in this laboratory frame and $\xi$ indicates the angle formed by the magnetic field at the tritium source measured counterclockwise from the local north.
This choice allows us to make use of symmetry properties of the spherical harmonics to perform the integration within the acceptance cone $\De\Om:\th\in[0,\th_0], \ph\in[0,2\pi]$ with ease.

Conventionally, near the endpoint the spectrum takes the form
\beq\label{BetaSpectrum}
\frac{d\Ga}{dT} = 3C_R\big[(\De T)^2-\ha m_\nu^2\BB],
\eeq
where $C_R$ is approximately constant and $\De T=T_0-T$ denotes the kinetic energy of the electron measured from the point $T_0$ where the spectrum would end in the absence of neutrino mass.
In the presence of Lorentz violation, the spectrum gets modified by the four components of the coefficient $\Aof$ in the form $T_0 \to T_0 + \de T$, with
\bea
\de T &=& \frac{1}{\De\Om}\int_{\De\Om}d\Om_{\nub} \, (\Aof)^\text{lab}
\nn\\
&=& \de T_\mathcal{C} + \de T_{\mathcal{A}_s} \sin\wT + \de T_{\mathcal{A}_c} \cos\wT,
\label{T(wT)}
\eea
which shows the sidereal-time dependence of this modification.
The amplitudes $\de T_\mathcal{C}$, $\de T_{\mathcal{A}_s}$, $\de T_{\mathcal{A}_c}$ are explicitly presented in Appendix \ref{App_T}.
The energy independence of $\de T$ allows a direct determination of the integrated spectrum 
\bea
\Ga(T) &=& \int_T^{T_\eff}\frac{d\Ga}{dT'}\,dT'
\nn\\
&=& C_R\big[(T_\eff-T)^3-\sF32 m_\nu^2(T_\eff-T)\BB],
\eea
where the effective null-mass endpoint energy $T_\eff = T_0 + \de T$ is a fit parameter that in the presence of Lorentz violation depends on the orientation of the experiment, location of the laboratory, and varies with sidereal time.
The use of MAC-E filters has been implemented in past by the Mainz \cite{Mainz} and Troitsk \cite{Troitsk} experiments, and unprecedent sensitivity will be achieved in KATRIN \cite{KATRIN}.
These experiments appear as ideal setups to search for the signals of Lorentz violation described here.

The study of Lorentz-violating neutrinos shows interesting features absent in other sectors.
In particular, the incorporation of Dirac and Majorana couplings as well as the implementation of the seesaw mechanism that suppresses the left-right handed mixing produces terms in the hamiltonian that appear as the product of the neutrino mass and a Majorana coefficient for CPT-even Lorentz violation \cite{KM2012}.
Some of these mass-induced coefficients $\C{2}{\eff}{jm}$ modify the neutrino mass measured as the parameter in the spectrum \eqref{BetaSpectrum} in the form $m^2_\nu\to m^2_\nu+\de m^2$, where the Lorentz-violating modification can be written in the form
\bea\label{m2(wT)}
\delta m^2 = m^2_\mathcal{C} + m^2_{\mathcal{A}_s}\,\sin\wT + m^2_{\mathcal{A}_c}\,\cos\wT,
\eea
to explicitly show the sidereal-time dependence of this parameter that mimics a neutrino mass.
The amplitudes $m^2_\mathcal{C}$, $m^2_{\mathcal{A}_s}$, $m^2_{\mathcal{A}_c}$ are explicitly presented in Appendix \ref{App_m2}.
This result shows that the experimental mass-squared parameter $m^2$ measured in the experiment includes the actual neutrino mass $m_\nu$ and a Lorentz-violating component that depends on the orientation and location of the laboratory as well as sidereal time.
Since there is no restriction on the sign of $\de m^2$, the coefficients $\C{2}{\eff}{jm}$ could even produce a negative $m^2$ without a tachyonic neutrino \cite{Chodos}.

\section{Conclusions}

In this article the low-energy signatures of Lorentz invariance violation in neutrinos in the context of the Standard-Model Extension
have been presented. 
The main focus is on a particular type of countershaded operator \cite{DKL} that is unobservable in neutrino oscillations and modifications to the neutrino velocity.
The main features that could arise in measurements of neutron decay as well as studies of the endpoint of beta decay are described.
Different experimental setups can be sensitive to the effects of this type of Lorentz violation, including a distortion of the entire beta spectrum in neutron decay, modifications to the measurement of the antineutrino-electron correlation in the decay of unpolarized neutrons, a correction to the electron-proton coincidence asymmetry in the decay of polarized neutrons, and a shift in the endpoint energy of the beta-decay spectrum.
A remark on the effects of a mass-induced coefficient is also presented in the context of tritium decay because these coefficients can mask the effects of the actual neutrino mass in novel ways.

Experimental signatures of the breakdown of Lorentz symmetry in the neutrino sector have been mostly explored using high-energy and interferometric phenomena; nonetheless, the high precision of low-energy experiments studying single beta decay can play a key complementary role in the search for deviations from exact Lorentz invariance.

\section*{Acknowledgments}
This work was supported in part by the Helmholtz Alliance for Astroparticle Physics (HAP) under Grant no. HA-301.

\newpage
\begin{widetext}
\appendix
\section{Sidereal amplitudes}

\subsection{Sidereal amplitudes for $\de B$}
\label{App_B}

The amplitudes for the sidereal decomposition of the Lorentz-violating experimental asymmetry defined in \eqref{B(wT)} are given by
\bea
\de B_\mathcal{C} &=& \sqrt{\frac{3}{\pi}} \,f(T)\cos\ch\, (\Aof)_{10},
\nn\\
\de B_{\mathcal{A}_s} &=& \sqrt{\frac{6}{\pi}} \,f(T)\sin\ch\,\Im(\Aof)_{11},
\nn\\
\de B_{\mathcal{A}_c} &=&-\sqrt{\frac{6}{\pi}} \,f(T)\sin\ch\,\Re(\Aof)_{11},
\eea
where $f(T)$ is a function of the electron's kinetic energy and other parameters.
Depending on the value of the factor $r$, the function $f(T)$ takes the form
\beq
f(T) =
\left\{ \begin{array}{ccc}
\dfrac{
(A\be-Br)(2-r^2)
-\Bz\Big[\Big(\frac{2r^2}{3}+\frac{2\be r}{3}-\be-2\Big)
+a\be\Big(1-\frac{2r^3}{5}\Big)\Big]
}{8-4r+a\be(r^2-2)}
&,& r<1 \\\\
\dfrac{(A\be-Br)+\Bz\Big[\frac{2\be}{3}+\frac{8r}{3}-\frac{6a\be}{5}\Big]}{4r-a\be}
&,& r>1 \end{array}\right..
\eeq

\subsection{Sidereal amplitudes for $\de T$}
\label{App_T}

The amplitudes for the sidereal decomposition of the Lorentz-violating shift of the endpoint energy defined in \eqref{T(wT)} are given by
\bea
\de T_\mathcal{C} &=& \arit - \sqrt{\frac{3}{4\pi}}\cos^2\frac{\th_0}{2}\sin\ch\,\cos\xi\,(\Aof)_{10},
\nn\\
\de T_{\mathcal{A}_s} &=&-\sqrt{\frac{3}{2\pi}}\cos^2\frac{\th_0}{2}
\big(\sin\xi\,\,\Re(\Aof)_{11}-\cos\xi\,\cos\ch\,\Im(\Aof)_{11}\big) ,
\nn\\
\de T_{\mathcal{A}_c} &=&-\sqrt{\frac{3}{2\pi}}\cos^2\frac{\th_0}{2}
\big(\sin\xi\,\,\Im(\Aof)_{11}+\cos\xi\,\cos\ch\,\Re(\Aof)_{11}\big) .
\eea

\subsection{Sidereal amplitudes for $\de m^2$}
\label{App_m2}

The amplitudes for the sidereal decomposition of the Lorentz-violating shift of the neutrino mass parameter defined in \eqref{m2(wT)} are given by
\bea
m^2_\mathcal{C} &=& 
\sqrt{\frac{3}{\pi}}\,\cos^2\frac{\th_0}{2}\,\sin\chi\,\cos\xi\, \C{2}{\eff}{10}  ,\nn\\
m^2_{\mathcal{A}_s} &=& 
\sqrt{\frac{6}{\pi}}\,\cos^2\frac{\th_0}{2}\,
\left[\sin\xi \,\, \Re\C{2}{\eff}{11}-\cos\chi \, \cos\xi \,\, \Im\C{2}{\eff}{11} \right],\nn\\
m^2_{\mathcal{A}_c} &=& 
\sqrt{\frac{6}{\pi}}\,\cos^2\frac{\th_0}{2}\,
\left[\sin\xi \,\, \Im\C{2}{\eff}{11} + \cos\chi \, \cos\xi \,\, \Re\C{2}{\eff}{11} \right].
\eea

\end{widetext}
\end{document}